\documentclass[aps,prl,reprint,10pt,twocolumn,nofootinbib,longbibliography,superscriptaddress]{revtex4-1}
\usepackage{amsmath}
\usepackage{amsfonts}
\usepackage{amssymb}
\usepackage{graphicx}
\usepackage{epstopdf}
\usepackage{color}
\usepackage{appendix}
\usepackage{csvsimple}
\usepackage{chngpage}
\usepackage{subcaption}
\usepackage{mathrsfs}
\usepackage[cm]{fullpage}
\usepackage[colorlinks,linkcolor=blue,citecolor=blue,urlcolor=blue]{hyperref}
\usepackage[table,xcdraw]{xcolor}
\usepackage[utf8]{inputenc}

\begin{document}

\title{Loop inflection-point inflation}
\date{\today}
\author{Konstantinos Dimopoulos}
\affiliation{Consortium for Fundamental Physics, Physics Department,Lancaster University, Lancaster LA1 4YB, United Kingdom}
\author{Charlotte Owen}
\affiliation{Consortium for Fundamental Physics, Physics Department,Lancaster University, Lancaster LA1 4YB, United Kingdom}
\author{Antonio Racioppi}
\affiliation{National Institute of Chemical Physics and Biophysics, R\"avala 10, 10143 Tallinn, Estonia}

\begin{abstract}
A novel inflection-point inflation model is analysed. The model considers a 
massless scalar field, whose self-coupling's running is stabilised by a 
non-renormalisable operator. The running is controlled by a fermion loop.
We find that successful inflation is possible for a natural value of the Yukawa coupling
\mbox{$y\simeq 4\times 10^{-4}$}. The necessary fine-tuning is only 
$\sim 10^{-6}$, which improves on the typical tuning of inflection-point 
inflation models, such as MSSM inflation. The model predicts a spectral 
index within the 1-$\sigma$ bound of the latest CMB observations, with a very small negative running, and negligible tensors (\mbox{$r\sim 10^{-(9-10)}$}). These results are largely independent of the order of the stabilising non-renormalisable operator. 
\end{abstract}

\maketitle

\nopagebreak

\section{Introduction} 
Cosmic inflation is an organic part of concordance cosmology. With a single 
stroke inflation addresses the fine-tuning problems of the hot big bang; namely
the horizon and flatness problems and also produces the primordial curvature 
perturbation, which seeds structure formation and is in excellent agreement 
with CMB observations \cite{planck}. According to the inflationary paradigm, 
the Universe undergoes inflation when dominated by the potential density of a 
scalar field, called the inflaton. However, the identity of the inflaton is as 
yet unknown.

The latest CMB observations suggest that the scalar potential of the inflaton 
features an inflationary plateau (e.g. see Ref~\cite{ours}). Numerous 
mechanisms have been put forward to generate such a plateau, involving exotic 
constructions in the context of elaborate, beyond-the-standard-model theories, 
such as superstrings. One such example is inflection-point inflation,
where the inflationary plateau is due to the interplay of opposing contributions
in the scalar potential, which (almost) cancel each other out generating a step 
on the otherwise steep potential wall. The original model was called A-term 
inflation, because it employed the A-term of a supersymmetric theory
\cite{A-term,juan}, or MSSM inflation, because it considered a flat direction 
in MSSM \cite{MSSM} as the inflaton. However, other models of inflection-point 
inflation have also been constructed \cite{inflection,inflection+}. Most of 
these also consider an elaborate setup in the context of supersymmetry, 
string theory or other extensions of the Standard Model. 

However, an advantage of the idea of inflation is that it does not have
to rely on exotic physics, in contrast to alternatives like the ekpyrotic 
scenario \cite{ekpyrotic} or string gas cosmology \cite{stringGas}. Indeed,
inflation may be realised simply within field theory in curved spacetime. It is also possible to achieve inflection-point inflation in this way. In this paper 
we explore such a possibility, where we exploit the loop corrections to the 
inflaton potential to generate the step-like plateau. This is similar to the works in Ref.~\cite{inflection+}. However, in Ref.~\cite{inflection+} the
authors consider a rather complicated running of the inflaton self-coupling, 
where many particles are contributing to it. We consider a simpler setup.

%
In previous works care was taken so that loop corrections do not spoil the 
stability of the potential \cite{loopmodels}. In contrast, here we consider a 
model in which the Coleman-Weinberg potential is unstable. Stability is recovered by introducing a Planck-suppressed effective operator. 

We use natural units where \mbox{$c=\hbar=1$} and \mbox{$8\pi G=m_P^{-2}$},
with \mbox{$m_P=2.43\times 10^{18}\,$GeV} being the reduced Planck mass.

\section{Coleman-Weinberg Potential}	
The general expression for the 1-loop potential is given by the 
Coleman-Weinberg (CW) result \cite{Coleman:1973jx}
	\begin{equation}
	V_\text{eff} = V +\sum_{i=1}^n 
\frac{g_i M_i^4 (\phi)}{64\pi^2} \ln\left(\frac{M_i^2 (\phi)}{\mu^2}\right) \, ,
	\label{eq:V:CW}
	\end{equation}
where $V$ is the tree-level potential, $\mu$ is the renormalisation scale and 
$M_i$ and $g_i$ are, respectively, the field dependent tree level mass and the 
number of intrinsic degrees of freedom of the particle-$i$ coupled with $\phi$. 
We assume a quartic tree-level potential for the inflaton field
	\begin{equation}
	V = \lambda \, \phi^4 \, ,
	\label{eq:V}
	\end{equation}
and that the dominant contribution in Eq.~(\ref{eq:V:CW}) is given by the 
Yukawa coupling $y$ between $\phi$ and a Weyl fermion\footnote{%
A similar computation can be performed also in the case of more fermionic 
degrees of freedom. However, since here we are not discussing the details of 
the fermion sector phenomenology, but just its contribution to the effective 
potential, we limit ourselves to the minimal setup.}. 
Therefore we can approximate Eq.~(\ref{eq:V:CW}) with 
	\begin{equation}
	V_{\text{eff}} (\phi) = 
\left[\lambda -\beta\ln\left(\frac{y^2 \phi^2}{\mu^2}\right) \right]\phi^4 \, ,
	\label{eq:V:CW:2}
	\end{equation}
where we used Eq.~(\ref{eq:V}) and $\beta =
y^4/32\pi^2$. We can improve the potential by inserting the running expression 
for $\lambda$. Since we assumed that the Yukawa coupling $y$ is the dominant 
contribution, a good approximation\footnote{\label{yfoot}
There is also a RGE for $y$ to be solved. In 
a minimal setup in which the Weyl fermion is only coupled to $\phi$, the beta 
function for such a coupling would behave as \mbox{$\beta_y \approx y^3$}. If 
\mbox{$y \ll 1$}, then the running of $y$ becomes negligible and $y$ can be 
safely treated as a constant.} 
for the RGE solution of $\lambda$ is
	\begin{equation}
\lambda (\mu) = \lambda (M) - 2 \beta  \log \left(\frac{\mu}{M}\right) \, ,
	\label{eq:run:lambda}	 
	\end{equation}
where $M$ is the scale at which we impose the boundary condition on the running 
of $\lambda$. Since we are interested in studying a configuration in which the 
CW potential is unstable, it is natural to pick\footnote{The choice is just a convenient parametrization. Even if we would assume $\lambda(M) \neq 0$, we can always find a new scale  \mbox{$M^* = M \exp(\frac{\lambda (M)}{2 \beta })$} at which $ \lambda (M^*) = 0$. Therefore the computations would then proceed in the same way from Eq. (\ref{eq:V:eff}) with simply $M^*$ in place of $M$.}
\mbox{$\lambda(M)=0$}.
Using this and inserting Eq.~(\ref{eq:run:lambda}) into Eq.~(\ref{eq:V:CW:2}) 
we get
	\begin{equation}
V_{\text{eff}} (\phi)= - \beta \ln\left(\frac{y^2 \phi^2}{M^2}\right) \phi^4\,.
	\label{eq:V:eff}
	\end{equation}

\section{Inflation Model with Inflection Point}
The potential in Eq.~(\ref{eq:V:eff}) is not stable because it
is unbounded from below. We assume that stability is ensured by 
the intervention of a non-renormalisable Planck-suppressed effective operator. 
Therefore let us consider the following inflaton potential
	\begin{equation}
V=-\beta\ln\left(\frac{y^2\phi^2}{M^2}\right)\phi^4+
\lambda_n\frac{\phi^{2n+4}}{m_P^{2n}},
\label{Vn}
	\end{equation}
where the first term is the 1-loop effective potential obtained in 
Eq.~(\ref{eq:V:eff}) and the second term is an effective non-renormalisable
operator, with \mbox{$\lambda_n\ll 1$} and \mbox{$n\geq 1$}. We consider only
the dominant non-renormalisable term, of order~$n$.

For the moment we choose \mbox{$n=1$} but later on we consider higher values 
of~$n$. For simplicity, we study the model where 
	\begin{equation}
	\frac{y^2}{M^2}= \frac{1}{m_P^2}\,. \label{eq:M}
	\end{equation}
If $y<1$ (required for pertubativity), it is possible to realise such a 
condition with sub-Planckian $M$. 

\emph{A priori}, $M$ and $y$ can take whatever possible value. 
However it is possible to reduce the parameters space, identifying a preferred region which is essentially described by Eq. (\ref{eq:M}). 
For example, assuming that our inflaton is not the Higgs boson of the
SM, it is reasonable to expect new physics to happen around the scale of grand unification (GUT-scale).
Therefore it is reasonable to consider $M \sim 10^{15-16}$ GeV.
In addition to that, the Yukawa coupling, $y$, generating the
loop correction must be small enough to preserve perturbativity, but on the other side, also big enough to give rise to relevant corrections.
Therefore a reasonable range for $y$ is\footnote{Indeed, we find $y=4 \times 10^{-4}$ (see conclusions), which is not that far from the expected range.}  around $10^{-(2-3)}$.
Combining the two expected regions for $M$ and $y$, we get that $y/M$ is
around $1/m_P$, therefore for the first analysis, in which we present a
new idea for inflection point models, it is enough to study the model
implementing Eq. (\ref{eq:M}). We will consider a broader range of $M$ and $y$ values in a future article.

Noting that
the slow-roll formalism is independent of the potential normalisation, 
we reparametrise the potential as
	\begin{equation}
	V = \beta\left[-\ln \left(\frac{\phi ^2}{m_P^2}\right)\phi ^4  + 
\alpha\frac{\phi ^6}{m_P^2} \right],
\label{V}
	\end{equation}
where \mbox{$\alpha=\lambda_1/\beta$}.
Such a potential has a flat inflection point at
	\begin{equation}
\phi_f = e^{1/4}m_P\quad{\rm and}\quad\alpha_f\equiv\frac{2}{3 \sqrt{e}} \,.
\label{phifalphaf}
	\end{equation}
To study the inflationary predictions for values of $\alpha$ around 
$\alpha_f$, we parametrise: 
\begin{equation}\label{alpha}
\alpha= (1+\delta)\alpha_f
\end{equation}
and use $\delta$ as a free parameter. 
Varying $\delta$ allows us to find the range of allowed slopes of the plateau 
around the flat inflection point.
Increasing $\delta$ increases the slope of 
the plateau. Decreasing $\delta$ to negative values introduces a local maximum.

There are two aspects to consider when constraining $\delta$. First, by contrasting the computed inflationary observables with the observations. Second, by 
ensuring that the necessary remaining e-folds of inflation since the 
cosmological scales exited the horizon, $N_*$, is not greater than the total e-folds of 
inflation, $N_{\rm tot}$. 
When the parameter space for $\delta$ 
is established we calculate predictions for the inflationary observables, namely
the spectral index of the scalar curvature perturbations, $n_s$, its running, 
\mbox{$n_s'\equiv\frac{d n_s}{d \ln k}$} and the tensor-to-scalar ratio, $r$.
%
%
%

\subsection{\boldmath Computing $N_*$}\label{sec:definingN}

First we must make clear the distinction between $N_{\mathrm{tot}}$
and $N_*$.
%
$N_{\rm tot}$ depends mainly on the initial conditions of the inflaton. 
We set the beginning of inflation to be determined by $\epsilon = 1$, 
where $\epsilon=-\dot H/H^2$ is the usual slow-roll parameter. 
For the e-folds of \textit{observable} inflation $N_*$, typically the reheating 
temperature has a large impact. However, our model does not need an in-depth 
investigation into reheating since in this model, after 
inflation, the field oscillates in a quartic minimum because of 
Eq.~(\ref{eq:V}) and also
\begin{equation}\label{lim}
\lim_{\phi\rightarrow 0}
\!\left[-\beta\ln\left(\frac{\phi^2}{m_P^2}\right)\phi^4\right]
=\frac12\beta\phi^4 \,.
\end{equation}
The average density of a scalar field coherently oscillating in a quartic 
potential scales as \mbox{$\rho_\phi\propto a^{-4}$} \cite{scaling}, just as the
density of a radiation dominated Universe. Hence, there is little distinction in
the expansion between inflaton oscillations and radiation domination after 
reheating, which means that $N_*$ is independent of the inflaton decay rate.

In this case we have
	\begin{equation}\label{eq:N*}
N_*=62.8 - \mathrm{ln}\Big(\frac{k}{a_0H_0}\Big) + 
\frac{1}{3}\mathrm{ln}\Big(\frac{g_*}{106.75}\Big) + 
\frac{1}{3}\mathrm{ln}\Big(\frac{V_{\mathrm{end}}^{1/4}}{10^{16}\mathrm{GeV}}\Big)
	\end{equation} 
where $k = 0.05\mathrm{Mpc^{-1}}$ is the pivot scale, $(a_0 H_0)^{-1}$ is the 
comoving Hubble radius today, $g_*$ is the effective number of relativistic 
degrees of freedom and $V_{\mathrm{end}}\equiv V(\phi_{\mathrm{end}})$, with `end' 
denoting the end of inflation. This simplifies when we take $g_* = 106.75$, 
corresponding to the standard model at high energies. Inputting the values 
of $k$ and $a_0H_0$ as well, gives
	\begin{equation}\label{eq:N*2}
N_* = 57.4 +
\frac{1}{3}\mathrm{ln}\Big(\frac{V_{\mathrm{end}}^{1/4}}{10^{16}\mathrm{GeV}}\Big)
\,.
	\end{equation}

\subsection{\boldmath Limits of $\delta$}\label{sec:limits}

$N_{\mathrm{tot}}$ can be calculated by integrating between the two values of 
$\phi$ that result in $\epsilon = 1$, marking the beginning and end points of 
slow roll inflation. If $N_* \simeq N_{\mathrm{tot}}$ we may need to investigate 
the initial conditions of $\phi$ to assess whether or not slow-roll does start 
at $\epsilon = 1$. This will depend on whether or not the inflaton is 
kinetically dominated when it reaches the plateau. 
Ensuring $N_{\mathrm{tot}}>N_*$ imposes a maximum value for $\delta$: 
%
\begin{equation}\label{eq:delta_constriant_N_tot}
\delta < 10^{-5.16} \,.
\end{equation}

\subsection{Inflationary Observables}

The spectral index and tensor-to-scalar ratio for this model are calculated for 
varying positive $\delta$ values and the parameter space satisfying the Planck 
results is presented in Table \ref{tab:n_s_r}, along with the values of $N_*$, 
$N_{\mathrm{tot}}$ and the running of the spectral index. Using the Planck 
$2-\sigma$ constraint of $n_s = 0.968 \pm 0.010$\cite{planck}, provides limits on $\delta$:
\begin{equation}\label{eq:delta_constraint_ns}
10^{-6.06} \leq \delta \leq 10^{-5.86} \,.
\end{equation}
It is clear that, for the region where the spectral index and tensor-to-scalar 
ratio values match observations, $\delta$ is within the constraint of 
Eq.~\eqref{eq:delta_constriant_N_tot} such that $N_* < N_{\mathrm{tot}}$ and we do not need to worry about initial conditions. 

\begin{figure}[b]
\centering
\includegraphics[width=1\linewidth]{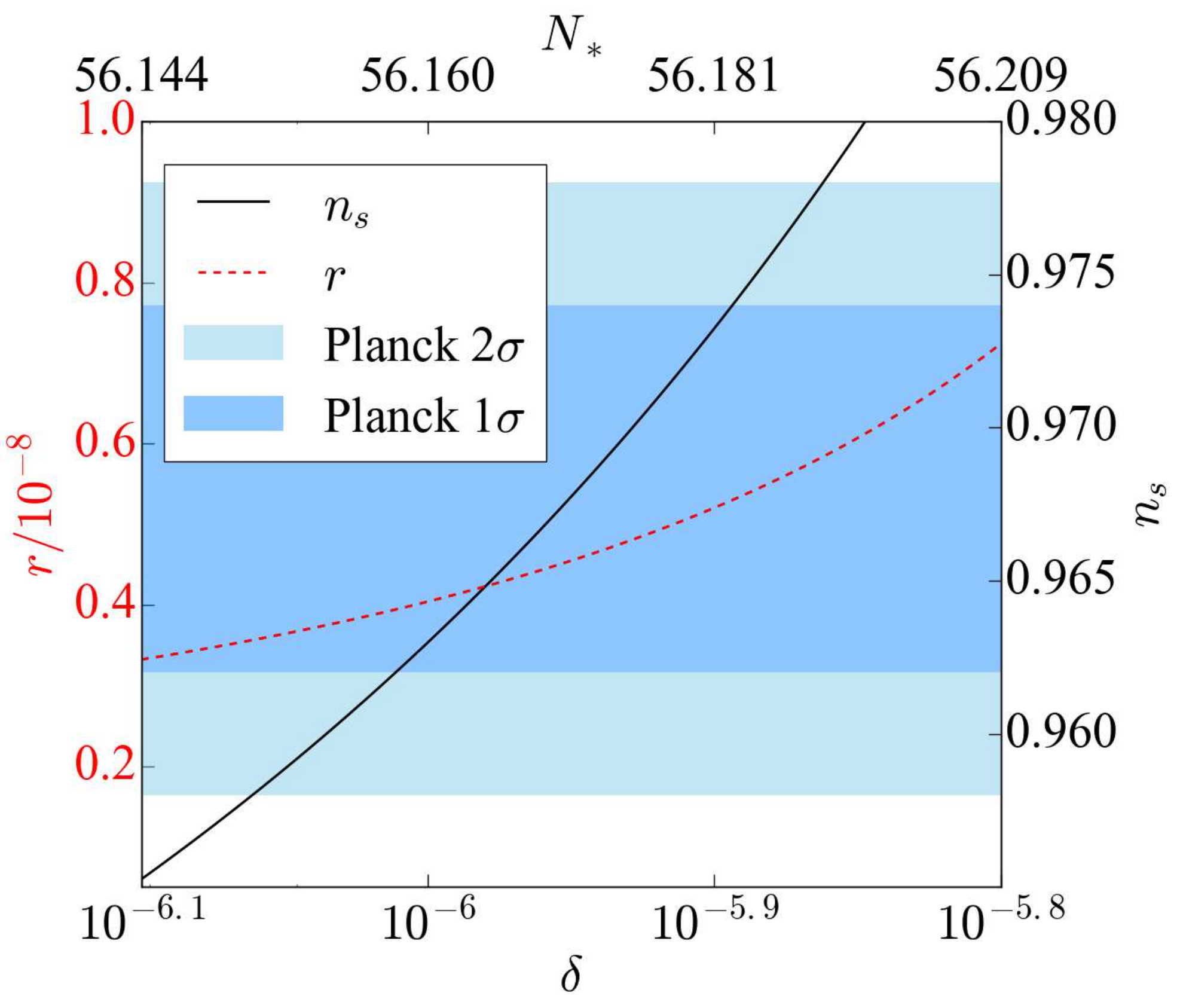}
\caption{Values of $\delta$ for which $n_s$ (solid black line) and $r$ (dashed
red line) fall within the Planck bounds for $n_s$ depicted with the shaded horizontal bands (light: 2-$\sigma$ and darkened: 1-$\sigma$). The top axis also shows the corresponding $N_*$ values for each $\delta$ value. (CMB only provides a weak bound on $r<0.1$\cite{planck})}
\label{fig:ns_r}
\end{figure}

The model's predictions for the inflationary observables are shown
in Table~\ref{tab:n_s_r} and Fig.~\ref{fig:ns_r}.

\begin{table}[h]
\setlength\extrarowheight{5pt}
\begin{center}
\begin{tabular}{|c|c|c|c|c|c|}\hline 
$\delta$   & $N_*$ & $N_{\mathrm{tot}}$ & $n_s$ & $r/10^{-9}$ & $n_s'/10^{-6}$ \\ 
				\hline 
$10^{-5.80}$ & 56.21 & 120.99 & 0.987 & 7.24 & -3.72 \\
				\hline 
$10^{-5.85}$ & 56.19 & 128.18 & 0.980 & 6.07 & -3.11 \\ 
				\hline
$10^{-5.90}$ & 56.18 & 135.79 & 0.973 & 5.21 & -2.67 \\ 
				\hline
$10^{-5.95}$ & 56.17 & 143.84 & 0.968 & 4.55 & -2.33 \\ 
				\hline
$10^{-6.00}$ & 56.16 & 152.38 & 0.963 & 4.04 & -2.07 \\
				\hline
$10^{-6.05}$ & 56.15 & 161.23 & 0.959 & 3.64 & -1.86 \\ 
				\hline
$10^{-6.10}$ & 56.14 & 171.00 & 0.955 & 3.32 & -1.70 \\
				\hline
\end{tabular} 
\end{center}
\caption{$\delta$ values producing $n_s$ within the Planck 2-$\sigma$ bounds.}
\label{tab:n_s_r}
\end{table}

\section{\boldmath Negative $\delta$ Values and Quantum Tunnelling}\label{sec:neg_d_QT}
When $\delta$ is negative, the potential develops a local minimum and maximum 
in place of a flat plateau. 
When the inflaton field tunnels through the local maximum, it may be in a 
position to slow-roll on the other side of the peak, hopefully for enough 
e-folds to generate $n_s$ and $r$ in accordance with observations. 
We calculated the number of slow-roll e-folds 
from the exit point of the quantum tunnelling and found that they are enough 
only when $\delta \geq -10^{-8}$ and in all cases 
$N_* = 56.08$. Even though this is a similar e-folding number to our results in the positive delta case, because the delta value is constrained to be a lot smaller it only results in $n_s = 0.928$ which is unacceptable because the 
spectrum is too red. 

\section{Higher-order non-renormalisable term}

\begin{figure}
	\centering
	\includegraphics[width=1.0\linewidth]{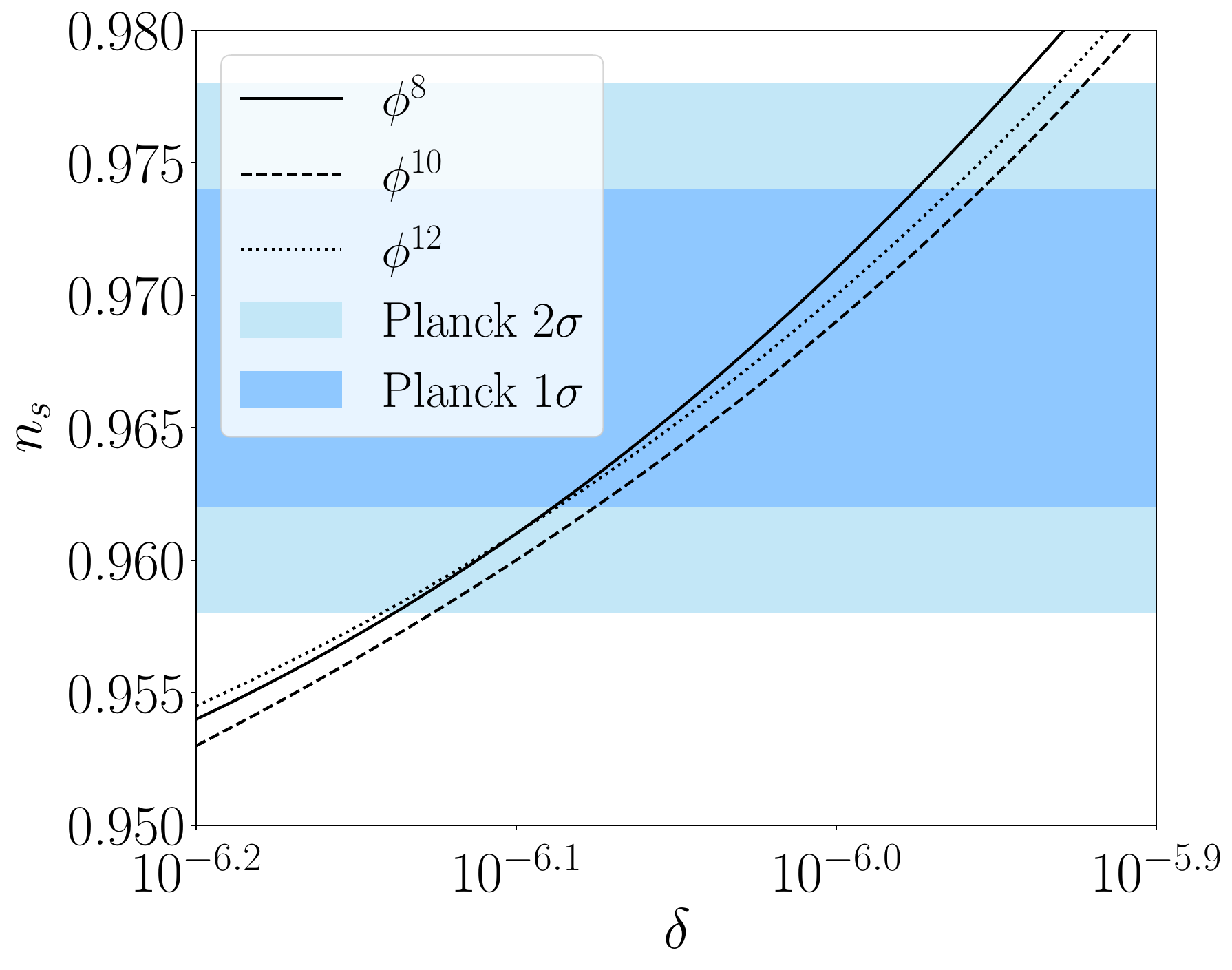}
	\caption{Values of $\delta$ for which $n_s$ falls within the Planck bounds depicted with the shaded horizontal bands \mbox{(light: 2-$\sigma$} and darkened: 1-$\sigma$) for varying orders of the non-renormalisable term, $n$. }
	\label{fig:dif_n_graph}
\end{figure}

Let us now consider higher values of $n$; the order of the non-renormalisable 
operator in Eq.~\eqref{Vn}. It is straightforward to check that our findings
in the \mbox{$n=1$} case are largely unchanged. As shown in 
Tables~\ref{phi8results}, \ref{phi10results} and \ref{phi12results}, 
for \mbox{$n=2,3,4$}, we 
find $n_s$ within the 2-$\sigma$ Planck bounds  only when 
\mbox{$10^{-6.2}<\delta<10^{-5.9}$}. We also find \mbox{$n_s'\sim - 10^{-6}$}, 
\mbox{$N_*\approx 56$}, \mbox{$N_{\rm tot}>2N_*$} and \mbox{$r\sim 10^{-(9-10)}$}. 
Our results show $r\simeq0$, with very small differences for varying $n$, and changes in $n_s$ for varying $n$ are at the $10^{-3}$ level. It should not be a surprise that our results are robust and largely independent 
of~$n$, as is clearly shown in Fig. \ref{fig:dif_n_graph}. By the time the cosmological scales leave the horizon the field has 
rolled passed the inflection point and the scalar potential is dominated by the 
CW-term in Eq.~\eqref{eq:V:eff},which is $n$ independent.

\begin{table}[h]
	\setlength\extrarowheight{3pt}
	\begin{center}
		\begin{tabular}{|c|c|c|c|c|}\hline
			$n$ & 1 & 2 & 3 & 4\\\hline
			$(\phi_f/m_P)^{2n}$ & 1.65 & 1.00 & 0.61 & 0.37 \\\hline
		\end{tabular}
		\caption{Values of $(\phi_f/m_P)^{2n}$ for $n\geq 1$.}
		\label{phif(n)}
	\end{center}
\end{table}

The value of the field at the inflection point~$\phi_f$ reduces somewhat for 
larger~$n$. 
Indeed, it is easy to show that
the generalisation of Eq.~\eqref{phifalphaf} for arbitrary $n$ is
\begin{equation}
\phi_f = e^{\frac{1}{2n}(1-\frac{n}{2})}m_P
\quad{\rm and}\quad
\alpha_f\equiv\frac{2 e^{\frac{n}{2}-1}}{n(n+2)} \,.
\label{phifalphaf+}
\end{equation}
The above suggest that \mbox{$(\phi_f/m_P)^{2n}=e^{1-\frac{n}{2}}$}, which means that
\mbox{$(\phi_f/m_P)^{2n}\lesssim 0.1$} for \mbox{$n\geq 4$} (see Table \ref{phif(n)}). 
This implies that, because \mbox{$\phi<\phi_f$} when the cosmological scales 
exit the horizon ($N_*<N_{\rm tot}/2$), we would expect high-order 
non-renormalisable terms to be suppressed when \mbox{$n>4$}. 
Thus, it is unlikely that the dominant, stabilising, non-renormalisable operator 
would correspond to \mbox{$n>4$}.


\begin{table}[h]
	\setlength\extrarowheight{5pt}
	\begin{center}
		\begin{tabular}{|c|c|c|c|c|c|}\hline
			$\delta$ & $N_*$ & $N_\mathrm{tot}$ & $n_s$ & $r / 10^{-10}$ & $n_s' /10^{-6}$ \\
			\hline
			$10^{-5.9}$ & 56.04 & 123.60 & 0.984 & $8.51$ & $-3.48$ \\\hline			
			$10^{-6.0}$ & 56.01 & 138.70 & 0.971 & $6.2$ & $-2.54$ \\\hline
			$10^{-6.1}$ & 55.99 & 155.64 & 0.961 & $4.88$ & $-1.99$ \\\hline
			$10^{-6.2}$ & 55.97 & 174.65 & 0.954 & $4.05$ & $-1.65$ \\\hline
		\end{tabular}
		\caption{Results for $\phi^8$ }
		\label{phi8results}
	\end{center}
\end{table}	
\begin{table}[h]
	\setlength\extrarowheight{5pt}
	\begin{center}
		\begin{tabular}{|c|c|c|c|c|c|}\hline
			$\delta$ & $N_*$ & $N_\mathrm{tot}$ & $n_s$ & $r / 10^{-10}$ & $n_s' /10^{-6}$ \\
			\hline
			$10^{-5.9}$ & 55.95 & 126.21 & 0.981 & $3.1$ & $-3.27$ \\\hline
			$10^{-6.0}$ & 55.93 & 141.81 & 0.969 & $2.30$ & $-2.42$ \\\hline
			$10^{-6.1}$ & 55.91 & 159.39 & 0.960 & $1.82$ & $-1.92$ \\\hline
			$10^{-6.2}$ & 55.90 & 179.23 & 0.953 & $1.52$ & $-1.60$ \\\hline	
		\end{tabular}
		\caption{Results for $\phi^{10}$ }
		\label{phi10results}
	\end{center}
\end{table}	
\begin{table}[h]
	\setlength\extrarowheight{5pt}
	\begin{center}
		\begin{tabular}{|c|c|c|c|c|c|}\hline
			$\delta$ & $N_*$ & $N_\mathrm{tot}$ & $n_s$ & $r / 10^{-10}$ & $n_s' /10^{-6}$ \\
			\hline
			$10^{-5.9}$ & 55.91 & 125.30 & 0.982 & $1.72$ & $-3.35$ \\\hline
			$10^{-6.0}$ & 55.88 & 142.71 & 0.970 & $1.29$ & $-2.51$ \\\hline
			$10^{-6.1}$ & 55.86 & 158.58 & 0.961 & $1.03$ & $-2.00$ \\\hline
			$10^{-6.2}$ & 55.85 & 172.96 & 0.954 & $0.87$ & $-1.68$ \\\hline
		\end{tabular}
		\caption{Results for $\phi^{12}$ }
		\label{phi12results}
	\end{center}
\end{table}	

\clearpage
\section{Inflationary scale and fine-tuning}
We determine the inflationary energy scale via the COBE constraint: 
	\begin{equation}
V^{1/4}=0.013\,r^{1/4} m_P\,.
	\end{equation}
As shown in Table~\ref{tab:n_s_r} and Tables~\ref{phi8results},
\ref{phi10results} and \ref{phi12results}, for \mbox{$\delta\sim 10^{-6}$}, we 
have \mbox{$r\sim 10^{-(9-10)}$}. Thus, the above 
suggests that \mbox{$V^{1/4}\sim 10^{14}\,$GeV}. Now, from Eq.~\eqref{V}, we have\footnote{Strictly speaking, Eq.~\eqref{V}  considers $n=1$. However,
observable inflation occurs after the inflaton field crosses the inflection point $\phi_f$, which means that the CW term dominates over the non-renormalisable term in Eq.~\eqref{V}. Thus, the order of the non-renormalisable term is not relevant here and
$V^{1/4}\sim\beta^{1/4}\phi$ for $n>1$ too.}
\mbox{$V^{1/4}\sim\beta^{1/4}\phi$}. 
Using the fact that 
\mbox{$\phi\sim\phi_f\sim m_P$} we obtain $\beta\sim 10^{-16}$. Because
\mbox{$\beta=y^4/32\pi^2$}, we find \mbox{$y \simeq 4\times10^{-4}$}, which is a 
very reasonable value for a Yukawa coupling and in agreement with the 
assumption \mbox{$y\ll 1$} (see Eq.~(\ref{eq:run:lambda}) and footnote~\ref{yfoot}). Through Eq.~\eqref{eq:M}, we then determine 
\mbox{$M=2\sqrt\pi(2\beta)^{1/4}m_P\simeq 10^{15}\,$GeV}; near the grand 
unification scale and sub-Planckian as expected.

Inflection-point inflation involves fine-tuning to attain the necessary 
inflationary plateau. In loop inflection-point inflation the tuning\footnote{This does not take into account the tuning required to satisfy Eq. (\ref{eq:M}). However such tuning is rather small since Eq. (\ref{eq:M}) is satisfied for quite natural values of the parameters.} is
\mbox{$\delta\sim 10^{-6}$}. This is exponentially better than the tuning 
corresponding to the horizon and flatness problems, resolution of which is 
one of the main motivations of inflation. For example, at the scale 
\mbox{$V^{1/4}\sim 10^{14}\,$GeV}, the deviation from flatness needs to be 
\mbox{$|\Omega-1|\lesssim 10^{-40}$}. Note also, that \mbox{$\delta\sim 10^{-6}$}
is much better than the level of tuning required in A-term/MSSM 
inflation~\cite{juan}. 

\begin{figure}
	\centering
	\includegraphics[width=1.0\linewidth]{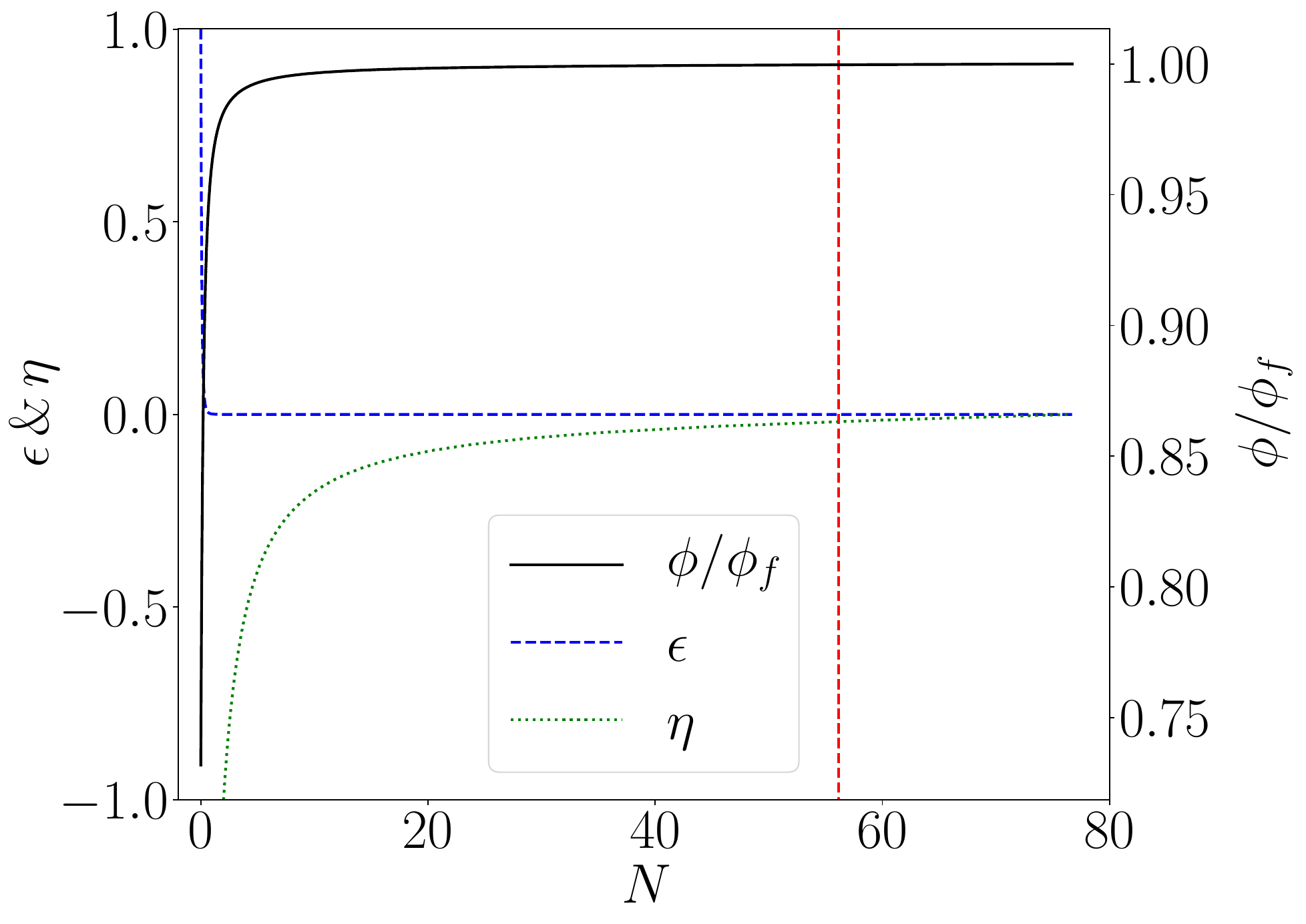}
	\caption{The values of the inflaton field $\phi$ (solid black line) and the slow-roll parameters $\epsilon$ (dashed blue line) and $\eta$ (dotted green line) with respect to the remaining e-folds of inflation $N$ are shown. The system progresses from right to left and inflation ends when \mbox{$N=0$}. The vertical (dashed red) line denotes $N_*$, which corresponds to the time when the cosmological scales exit the horizon during inflation. The inflaton is taken to roll from the inflection point at $\phi_f$ with negligible initial kinetic density, such that the slow-roll attractor is immediately assumed. As shown, $\epsilon$ is kept exponentially small during inflation. Inflation ends when \mbox{$\epsilon=1$}, with $|\eta|$ becoming large in the last e-fold of inflation, corresponding to a substantial variation of $\epsilon$, that ends inflation.}
	\label{fig:newgraph}
\end{figure}

It is important to note here that the assumption of slow-roll is not always
justified in inflection-point inflation models. This is because the potential
near the inflection-point is so flat that the system may depart from slow-roll 
and temporarily engage into so-called ultra-slow-roll (USR) inflation \cite{usr}\footnote{We would like to thank C.~Germani for pointing this out.}. This
can have profound implications on the calculation of inflationary observables
and may invalidate our findings (as well as those of most of the 
inflection-point literature). However, this danger can by averted if we assume 
that the inflaton lies initially near the inflection point with small enough 
kinetic density. In Ref.~\cite{USR} it is shown that, when the original kinetic 
density satisfies the bound \mbox{$\rho_{\rm kin}\leq(V'm_P)^2/6V$} at the 
inflection point, then slow-roll inflation begins immediately and all our 
findings are reliable. In our model \mbox{$V\sim\beta\,m_P^4$} and it can be 
easily shown that \mbox{$V'(\phi_f)\sim\beta\delta\,m_P^3$}. This means that, if
the inflaton starts near $\phi_f$ with kinetic density
\mbox{$\rho_{\rm kin}\lesssim\beta\delta^2 m_P^4$} then slow-roll inflation begins
immediately and our findings are fine. Putting in the numbers, we find
\mbox{$\rho_{\rm kin}^{1/4}\lesssim10^{11}\,$GeV}, i.e. a factor of $10^3$ smaller 
than the energy scale of inflation. In Fig. \ref{fig:newgraph} we track the evolution of the inflaton field and slow-roll parameters during inflation to demonstrate the avoidance of USR for negligible initial kinetic energy densities at the inflection point. Note that the issue of the initial 
conditions of inflation is academic because of the no-hair theorem, which 
demonstrates that memory of the initial conditions is lost once the inflationary
attractor is attained.

\section{Reheating}

A theory with \mbox{$V\sim\beta\phi^4$}, when \mbox{$V''>H^2$} leads to coherent 
oscillations in a quartic potential, whose density scales as radiation 
\mbox{$\rho\propto a^{-4}$} \cite{turner}. This means that the amplitude of the oscillations decreases as \mbox{$\phi\propto 1/a$}. These oscillations 
correspond to particles of mass \mbox{$m\sim\sqrt\beta\,\phi\propto 1/a$},
which is redshifted similarly to radiation particles \cite{Shtanov:1994ce}.

The decay rate of the inflaton particles to the Weyl fermions they couple to is
\mbox{$\Gamma=y^2m/8\pi$}. This means that, after inflation and during the 
oscillations, \mbox{$\Gamma/H\propto a$}, since \mbox{$H\propto a^{-2}$}. Thus,
because $\Gamma/H$ increases in time, there will be a moment when 
\mbox{$\Gamma\sim H$}and the decay becomes efficient and it leads to reheating.
Here we assume that the Weyl fermions produced by the inflaton decay are
coupled to SM particles such that, once formed, they promptly decay into the 
radiation bath of the hot big bang. 

Therefore, reheating occurs when
\begin{equation}
\frac{a_{\rm reh}}{a_{\rm end}}\sim\frac{H_{\rm end}}{\Gamma_{\rm end}}\sim
\frac{1}{\sqrt\beta}\frac{8\pi}{y^2}\frac{H_{\rm end}}{\phi_{\rm end}}\,,
\end{equation}
where `reh' denotes the moment of reheating.
Using that the amplitude of the oscillations decreases as 
\mbox{$\phi\propto 1/a$}, the above gives
\begin{equation}
\phi_{\rm reh}\sim\frac{y^2\sqrt\beta}{8\pi}\frac{\phi_{\rm end}^2}{H_{\rm end}}\,.
\label{phireh}
\end{equation}

For the density of the oscillating condensate we have 
\begin{equation}
\rho_{\rm reh}=\rho_{\rm end}\left(\frac{a_{\rm end}}{a_{\rm reh}}\right)^4\sim
9\beta\left(\frac{y^2}{8\pi}\right)^4m_P^4\,,
\end{equation}
where we used \mbox{$\rho_{\rm end}\sim V_{\rm end}\sim\beta\phi_{\rm end}^4$} and
\mbox{$H_{\rm end}^2=\rho_{\rm end}/3m_P^2$}. 

For the radiation bath we have \mbox{$\rho_{\rm reh}=(\pi^2/30)g_* T_{\rm reh}^4$},
where $g_*$ is the effective relativistic degrees of freedom and $T_{\rm reh}$ is 
the reheating temperature. Using this, the above equation suggests
\begin{equation}
T_{\rm reh}\sim\left(\frac{270}{\pi^2g_*}\right)^{1/4}
\frac{y^2\beta^{1/4}}{8\pi}\,m_P\sim 0.03\,y^2\beta^{1/4}m_P\,,
\end{equation}
where we considered that \mbox{$g_*={\cal O}(100)$}.%
\footnote{This is Eq.~(108) of Ref.~\cite{Shtanov:1994ce}.} Putting the numbers we 
obtained: \mbox{$y\simeq 4\times 10^{-4}$} and \mbox{$\beta\sim 10^{-16}$}, we get
\mbox{$T_{\rm reh}\sim 10^6\,$GeV}, which is comfortably higher than the 
temperature at BBN ($\sim 1\,$MeV) but low enough to avoid the generation of 
dangerous relics (e.g. gravitinos).

Now suppose that there is also a quadratic term in the scalar potential such 
that the inflaton has a bare mass $m_0$ and 
\mbox{$V\sim m_0^2\phi^2+\beta\phi^4$}. In order not to influence inflation, the
quadratic term must remain negligible during inflation. This means 
\mbox{$m_0^2<\beta\phi^2$}. Using \mbox{$\beta\sim 10^{-16}$} and 
\mbox{$\phi\sim\phi_f\sim m_P$}, we find the bound \mbox{$m_0<10^{10}\,$GeV}.

In order not to influence reheating the bound on $m_0$ is much more stringent,
because we need the quadratic term in the potential to remain subdominant 
until the decay of the inflaton condensate, that is we need 
\mbox{$m_0^2<\beta\phi_{\rm reh}^2$}. In view of Eq.~(\ref{phireh}), we get
\begin{equation}
m_0<\sqrt{3\beta}\,\frac{y^2}{8\pi}\,m_P\,,
\end{equation}
where we also considered that \mbox{$H_{\rm end}^2\simeq V_{\rm end}/3m_P^2$} and 
\mbox{$V_{\rm end}\sim\beta\phi_{\rm end}^4$}. Putting the numbers in, we obtain
\mbox{$m_0<300\,$GeV} or so. This is a bit tight but it also means that 
if \mbox{$m_0\sim 1\,$TeV}, the influence on the value of $N_*$ 
would be of the order 
\mbox{$\Delta N_*\simeq\frac16\ln(m_0^2-\beta\phi_{\rm reh}^2)<1$}, which 
would have minimal impact on our results, 
while a TeV-scale scalar particle might be 
observable in the LHC in the near future.

\section{Conclusions}
To conclude, we have studied a simple but elegant inflation model, where the 
inflationary plateau is generated through the running of the self-coupling 
of a massless scalar field, stabilised by a non-renormalisable operator. We 
have found that the model accounts for observations with mild tuning of the 
order $\sim 10^{-6}$ and a natural value of the Yukawa coupling 
\mbox{$y\simeq 4\times 10^{-4}$}. In particular, the model can result in the 
spectral index of the scalar curvature perturbation within the 1-$\sigma$ bound 
of the latest CMB observations, while producing negligible tensors 
(\mbox{$r\sim 10^{-(9-10)}$}). The inflationary energy scale is 
\mbox{$V\sim 10^{14}\,$GeV}; much higher that A-term/MSSM inflation (hence, the 
tuning is less). We also studied perturbative reheating in our model and obtained a reasonable reheating temperature $T_{\rm reh}\sim 10^6\,$GeV. Non-perturbative effects might enhance the efficiency of reheating. Our setup is minimal and does not require exotic physics apart 
from the non-renormalisable term.

\paragraph{Acknowledgements}
KD is supported (in part) by the Lancaster-Manchester-Sheffield Consortium for 
Fundamental Physics under STFC grant: ST/L000520/1. CO is supported by the FST 
of Lancaster University. AR is supported by the Estonian Research Council grants IUT23-6, PUT1026 and by the ERDF Centre of Excellence project TK133.

\end{document}